\documentclass[aps,nofootinbib,twocolumn,floatfix,showpacs,preprintnumbers,prd]{revtex4}
\usepackage{graphicx, epsfig, bm, amsmath}
\usepackage{color}
\usepackage{hyperref}
\hypersetup{pdftitle="Super-exponential cutoff",pdfdisplaydoctitle}
\def\mnras{Mon.~Not.~Roy.~Astron.~Soc.}
\usepackage{ wasysym }

\begin{document}

%%%%%%%%%%%%%%%%%%%%%%%%%%%%%%%%%%

%\addtolength{\parskip}{-0.5mm}

\definecolor{darkgreen}{cmyk}{0.85,0.2,1.00,0.2}
\newcommand{\ab}[1]{\textcolor{red}{[{\bf AB}: #1]}}
\newcommand{\js}[1]{\textcolor{blue}{[{\bf JS}: #1]}}
\newcommand{\gB}{g_B}
\newcommand{\WP}{W}
\newcommand{\XP}{X}
\newcommand{\B}{B^{\rm Bulk}}
\newcommand{\tm}{m}
\newcommand{\sh}{M}

%\setlength{\parskip}{0pt}
%%%%%%%%%%%%%%%%%%%%%%%%%%%%%%%%%%%%%%%%%%%

\title{Super-exponential cutoff as a probe of annihilating dark matter}
\author{Alexander V. Belikov$ ^1$ and Joseph Silk$1^{,2,3}$}
%\affiliation{Institut d'Astrophysique de Paris, UMR 7095 CNRS, Universit\'e Pierre et Marie Curie, 98 bis Boulevard Arago, Paris 75014, France}
%\author{Joseph Silk}
\affiliation{$^1$Institut d'Astrophysique de Paris, UMR 7095 CNRS, Universit\'e Pierre et Marie Curie, 98 bis Boulevard Arago, Paris 75014, France,}
\affiliation{$^2$Department of Physics and Astronomy, 3701 San Martin Drive, The Johns Hopkins University, Baltimore MD 21218, USA,} 
\affiliation{$^3$BIPAC, 1 Keble Road, University of Oxford, Oxford OX1 3RH UK}
\begin{abstract}
We investigate the possibility of fitting the putative spectrum of gamma-rays from annihilating dark matter at the Galactic Center by a super-exponential template. By generating and analyzing mock data for several ongoing and future experiments (HESS, CTA and DMA), we find that the preference for a super-exponential template (over an exponential) is especially strong for dark matter annihilating to $\tau^+\tau^-$ and $\mu^+\mu^-$ relative to other annihilation channels.
\end{abstract}


\maketitle

%=================================================================

The discovery of new physics often relies on our ability to distinguish signal from background. 
Sometimes, however, the 'less interesting' background turns out to be no better known than the signal that we are seeking. When both signals have similar magnitudes, one could make use of the spectral differences between the signal and the background if they were sufficiently strong.

The study of spectral features of astrophysical signals 
as the putative spectrum of annihilating dark matter has recently attracted much attention. The most distinct signature could be provided by a line feature 
\cite{1986PhRvL..56..263S, Bringmann:2011ye}, which may have been observed in the Galactic Center region at 130 GeV \cite{Bringmann:2012vr} 
but this remains controversial
 \cite{Whiteson:2013cs}. Less dramatic, but possibly more realistic, spectral features that are hitherto unobserved are the so-called cut-off features \cite{Bringmann:2011ye}, and box-shaped spectra features \cite{Ibarra:2012dw}. 

Here we study the spectral properties of a hypothetical gamma-ray signal of dark matter annihilating in the Galactic Center region, and demonstrate that a simulated spectrum of dark matter annihilation to hard channels (such as $\tau^+\tau^-$ and $\mu^+\mu^-$) can be better fit by the so-called 'super-exponential' fitting function rather than by the commonly adopted exponential fit. 

As a template dark matter annihilation model where the signal is potentially strong, we adopt the dark matter density spike that forms around a central black hole in a dark matter halo or sub halo \cite{1999PhRvL..83.1719G}. This is visible as a point source, but similar arguments would apply to a diffuse dark matter concentration.

The point source at the Galactic Center, associated with the central black hole Sgr A*, has been identified in radio, infrared and X-ray bands, as well as in gamma-rays by HESS \cite{Aharonian:2004wa}, where
 it appears as a point source within the angular resolution of the instrument ($\Delta\Omega \simeq 10^{-5}$). The HESS Collaboration identified a possible exponential cut-off in the spectrum of the Galactic Center source at about 15 TeV \cite{Aharonian:2009zk}, which could also be explained by a broken power-law.
The gamma-ray emission from source HESS J1745-290 and its high energy cutoff was explained by astrophysical mechanisms in terms of acceleration and diffusion of cosmic rays \cite{Aharonian:2005ti}, and alternatively in terms of annihilating dark matter \cite{Profumo:2005xd, Belikov:2012ty}. The next generation ground-based telescope CTA will provide superior spectral and angular resolution \cite{Becherini:2012iy}. For completeness in our analysis we also invoke the concept of the more futuristic experiment, the Dark Matter Array (DMA) \cite{Bergstrom:2010gh}. With such experiments in perspective, we argue below that improved spectral measurement of the putative TeV cut-off could discriminate between competing models. 
We describe the technique used to simulate the data and present the results of fits to the simulated data by exponential and super-exponential functions.

The mean number of events in the energy bin $E_i$, given the flux $\phi(E)$ registered by an experiment with an effective collecting area $A(E)$ during time $\Delta t$, is estimated as
$$
N_i = \frac{\Delta t}{\sigma_i\sqrt{2\pi }} \int_{\Delta E_i} dE \int dE' A_{eff} (E')\phi(E') e^{-\frac{(E'-E)^2}{2\sigma_i^2}},
$$
where $\Delta E_i$ is the width of the energy bin and $\sigma_i$ is the variance of energy, related to energy resolution of the instrument ($\sigma_i = E_i \delta_{res} / \sqrt{8\ln2}$). For the sake of simplicity, we take the effective area $A_{eff} (E) = A_{eff} (E = 1 \mbox{TeV})$ to be constant. Once we have the mean number of events, we generate 100 realizations of the observations for a given bin and calculate the statistical error assuming a Poisson distribution. The total error includes the statistical as well as instrument specific systematic error $\delta_{syst}$. The values of energy resolution, typical observation times, effective areas and systematic errors for HESS, CTA and DMA are summarized in Table \ref{tab:Stats}.

\begin{table}[h]
\centering
\begin{tabular}{c|c|c|c|c|c}
\hline
 & $\Delta t$ [hours] & $A_{eff}$ [$10^6 \,\mbox{m}^2$] & $\delta_{res}$ & $\delta_{syst}$ & $f$\\ 
\hline
HESS & 50 & 0.18 & 15\% & 10\% & 0.1\\
CTA & 100 & 2.3 & 9\% & 5\% & 0.01\\
DMA & 5000 & 23 & 7\% & 1\% & 0.001\\
\hline
\end{tabular}
\caption{Benchmark parameters of current and future air Cherenkov telescope observations of the Galactic Center.}
\label{tab:Stats}
\end{table}

\vskip -8pt 

We simulate the flux measured from the Galactic Center region as a sum of the dark matter annihilation spectrum plus a power-law-like background $\phi(E) = \phi_\chi(E) + \phi_{bg}(E)$. While ground-based Cherenkov telescopes cannot distinguish gamma-rays from electrons and positrons, we assume that the major part of the electron and positron background can be excluded from the Galactic Center region due to the fact 
that the observed flux of electrons and positrons is approximately globally isotropic (perhaps up to a small dipole anisotropy \cite{Hooper:2008kg}).

If dark matter consists of particles of mass $M_\chi$ that annihilate with a cross section $\langle \sigma v\rangle$, the flux of gamma rays from a dark matter halo with the density distribution described by $\rho(l, \psi, \phi)$ is given by 
$$
\phi_\chi(E) = \frac{\langle \sigma v\rangle}{ 8 \pi M^2_\chi} \frac{dN_\gamma}{dE} \int \rho^2(l, \psi, \phi) dl d\Omega,
$$
where $\frac{dN_\gamma}{dE}$ is the spectral density of $\gamma$'s per annihilation. 

N-body simulations find that a cusp is formed at the center of a dark matter halo. The slope of the cusp was found to range from $\rho \propto r^{-1}$ to $\rho \propto r^{-1.5}$ \cite{Navarro:1995iw, Moore:1999gc}. 
Cores were observed in dwarf galaxies \cite{Moore:1994yx}.
In case a black hole resides at the center of a halo, the density profile of the halo is expected to contract in its vicinity. 
The slope of the spike is derived in terms of the inner slope $\gamma$ as $\gamma_{sp} = 2 + 1/({4-\gamma})$ in the simplest adiabatic contraction model \cite{1999PhRvL..83.1719G}. The enhancement of the dark matter annihilation signal over the non-contracted profile can reach a factor of 100 in the innermost region of the Galaxy. 

We normalize our dark matter profile to the dark matter density in the vicinity of the Solar system, taken to be $\rho_{\odot} (R_\odot = 8.5\, \mbox{kpc}) = 0.3\,{\rm GeV}\,{\rm cm}^{-3}$
\cite{Bovy:2012tw}.
We use a generalized NFW profile with $\gamma = 1.42$ or equivalently take the flux to be boosted by a factor of $b=400$ over the standard NFW density profile with $\gamma = 1.0$ (consistent with \cite{Belikov:2012ty}) :
$$
\phi_\chi(E) = 1.1 \times 10^{-7} \frac{\langle \sigma v\rangle}{\langle \sigma v\rangle_{th}}
\left(\frac{M_\chi}{M_0}\right)^{-2} \frac{dN}{dE}(E) \, \mbox{cm}^{-2} \mbox{s}^{-1},
$$
with $\langle \sigma v\rangle_{th} = 3\times 10^{-26} \mbox{cm}^{-3}\mbox{s}^{-1}$ and $M_0 = 100 \,\mbox{GeV}$.
This profile and/or boost is equivalent to adopting a density spike, for which the conventional assumption is $\gamma^\prime =7/3$ within $\sim 0.1$ pc \cite{1999PhRvL..83.1719G}.

The astrophysical diffuse background gamma-ray flux is modeled by a power law $\phi_{bg}(E) = A_{bg} E^{-\Gamma_{bg}}$ with $\Gamma_{bg} = 2.7$ 
 \cite{Bergstrom:1997fj}.
Estimates for the spectral index of the diffuse gamma-ray flux vary from $2.4$ for the blazar component \cite{Venters:2009sd} to $2.75$ \cite{Fields:2010bw} for the component from star-forming galaxies.
The measurement of the spectral index of the diffuse gamma-ray flux by Fermi in the range 1 to 50 GeV provided $\Gamma_s = 2.40 \pm 0.07$ \cite{Abdo:2010nz}, while a certain population of sources, such as radio quasars, are characterized by a steeper spectral index $\Gamma_{rq} = 2.48 \pm 0.02$ \cite{Collaboration:2010gqa}.

An exponential cutoff is a template often used for describing rapidly decreasing functions. The form $\phi(E) = A E^{-\Gamma} e^{-(E/E_{cut})}$ has been used to describe the gamma-ray spectrum of the source at the Galactic Center \cite{Aharonian:2006wh}. The continuum spectrum of annihilation of dark matter into photons for most annihilation channels has a pronounced cutoff at the mass of the dark matter particle. It is tempting to model such an abrupt cutoff with functions that decay more rapidly than an exponential function. A 'super-exponential' function $\phi(E) = A E^{-\Gamma} e^{-(E/E_{cut})^\beta}$ is a simple generalization of an exponential function.

\begin{figure*}[h]
\centering
\includegraphics[width=0.9\linewidth,clip=]{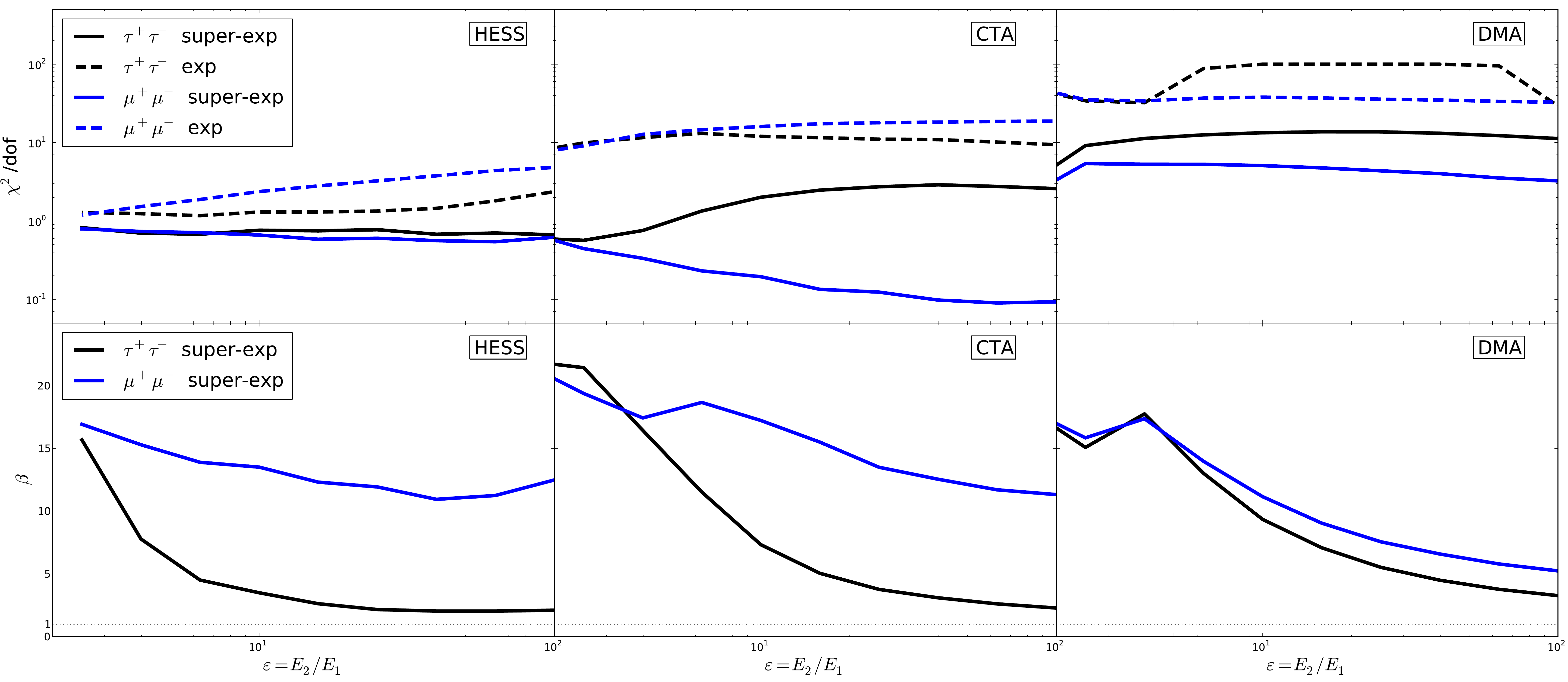}
\caption{
Top: averaged $\chi^2/\mbox{dof}$ of the best fits as a function of $\varepsilon = E_2/ E_1$: 
exponential function, dashed lines;
super-exponential function, solid lines. Black lines correspond to $\tau^+ \tau^-$ annihilation channel, and blue lines to $\mu^+ \mu^-$. 
For normalization of background, see Table \ref{tab:Stats}.
Bottom: the parameter $\beta$ best fits to simulated datasets of the dark matter signal and background averaged over realizations as a function of $\bar E$:
black solid lines, $\tau^+ \tau^-$ annihilation channel; blue solid lines, $\mu^+ \mu^-$ annihilation channel. For all these plots $\bar E = \varepsilon^{-0.25} M_\chi$.
From left to right : HESS, CTA, DMA.
}
\label{fig:FunctionsOfEps}
\end{figure*}

We generate datasets that are potentially observeable by HESS, CTA and DMA for $\tau^+\tau^-$, $\mu^+\mu^-$, $W^+ W^-$ and $b \bar b$ channels for dark matter with $M_\chi = 4$ TeV using the PPPC4DMID code \cite{Cirelli:2010xx}. After a few tests we found that for our choices of experimental parameters, the $\tau^+\tau^-$ and $\mu^+\mu^-$ annihilation channels benefit most from using the super-exponential spectral templates, and we decided to work specifically with these channels.
Using the technique of 'scalable' windows, we bin the data in $N_b = 20$ logarithmic bins between $E_a$ and $E_b$, for a range of $\varepsilon = E_b/E_a$. As found earlier, it is propitious to keep the geometric mean ($\bar E = \sqrt{E_b E_a}$) in a fraction of the dark matter mass $\bar E = \varepsilon^{-0.25} M_\chi$ \cite{Bringmann:2011ye}. To parametrize the background flux against the flux due to dark matter annihilation we use the fraction of energy contained in the background $f = \int E \phi_{bg}(E) dE / \int E \phi_{\chi}(E) dE$, where the integration is performed over the interval $[E_a, E_b]$. 
For each $\varepsilon$ in the interval [1,100] and each dataset, we find the best-fit parameters of the exponential and the super-exponential fits, and calculate the goodness-of-fit parameter $\chi^2$. The average values of $\chi^2/\mbox{dof}$ and $\beta$ are presented in
Fig. \ref{fig:FunctionsOfEps}. 

Since the super-exponential function has one extra parameter $\beta$ over the exponential function, we expect that $\chi^2$ values for the former are at least as good as for the latter, which we find for the specified values of parameter $f$ (Table \ref{tab:Stats}) for the whole range of $\varepsilon$.
For HESS, the fits with the super-exponential function prove to be good in the whole range of varying $\varepsilon$ for $\mu^+\mu^-$ and $\tau^+\tau^-$ annihilation channels, while for CTA, the $\mu^+\mu^-$ channel fits well across the whole range, and the $\tau^+\tau^-$ annihilation channel can be well fit with a super-exponential only in the range $\varepsilon < 10$.
However, the DMA fits with the super-exponential function are not as good for both channels ($\chi^2/dof \approx 10$). This is due to the fact that with such overwhelming statistics and systematics, a simple super-exponential will no longer be a sufficiently good template to describe the sum of the signal and the power-law background.
From Figure \ref{fig:FunctionsOfEps} one can see that the average exponent $\beta$ of the best fits is a nearly monotonically decreasing function of $\varepsilon$. For CTA and DMA and $\varepsilon > 3$ and across the whole range for HESS, the super-exponential fit to the $\mu^+\mu^-$ annihilation channel requires a greater $\beta$ to fit the observed flux. The variance of $\chi^2$ does not exceed 30\% for super-exponential fits to HESS spectra and is much less than that for the rest of the scenarios. With the increase of $\Gamma$ in the range [2.4,2.7], the mean $\chi^2$ of exponential fits increases and that of the super-exponential plots decreases, the maximum variation does not exceed 6\%. The mean $\beta$ of best fits increases with the increase of $\Gamma$ with the maximum variation below 1\%.

We note that the super-exponential fits for $W^+W^-$ and $b\bar b$ channels for CTA and DMA experiments to the values of $\chi^2$ and $\beta$ as functions of $\varepsilon$ are very similar to the $\mu^+\mu^-$ and $\tau^+\tau^-$ annihilation channels. However we find that for the HESS experiment the corresponding values of $\beta$ are not consistently greater than unity. This might be due to the fact that there is less intensity at the higher energies (close to $M_\chi$) for these channels.

\begin{figure*}[h]
\centering
\includegraphics[width=0.9\linewidth,clip=]{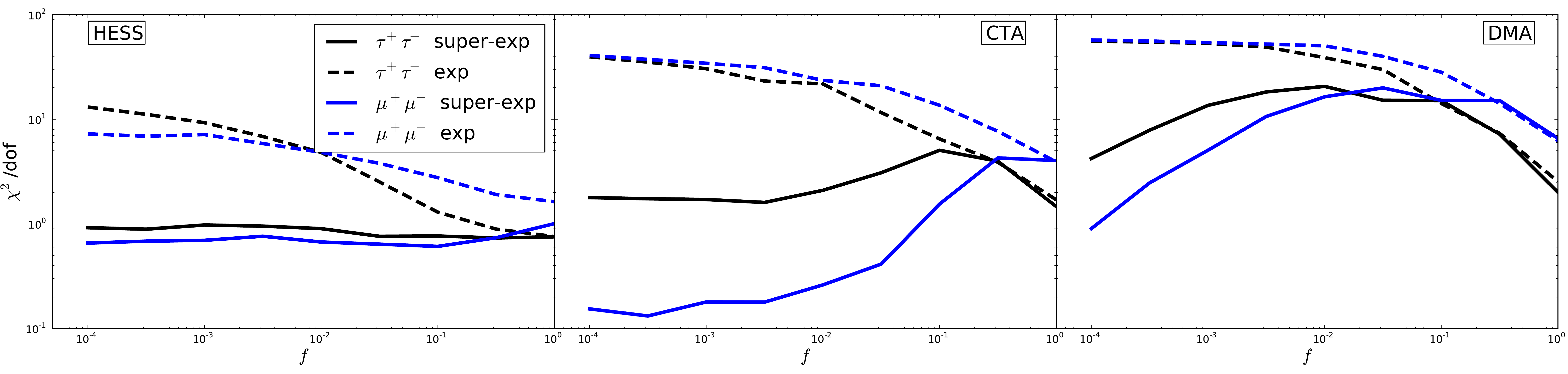}
\caption{
Averaged $\chi^2/\mbox{dof}$ of best fits as a function of $f$, the fraction of energy in the background: 
exponential function,
dashed lines; super-exponential function, solid lines. Black lines correspond to $\tau^+ \tau^-$ annihilation channel, blue lines - to $\mu^+ \mu^-$. Here $\varepsilon = 10$, $\bar E = \varepsilon^{-0.25} M_\chi$. From left to right : HESS, CTA, DMA.
}
\label{fig:chi2AsFunctionOfApl}
\end{figure*}
\begin{figure*}[h]
\centering
\begin{tabular}{ccc}
\includegraphics[width=0.3\linewidth,clip=]{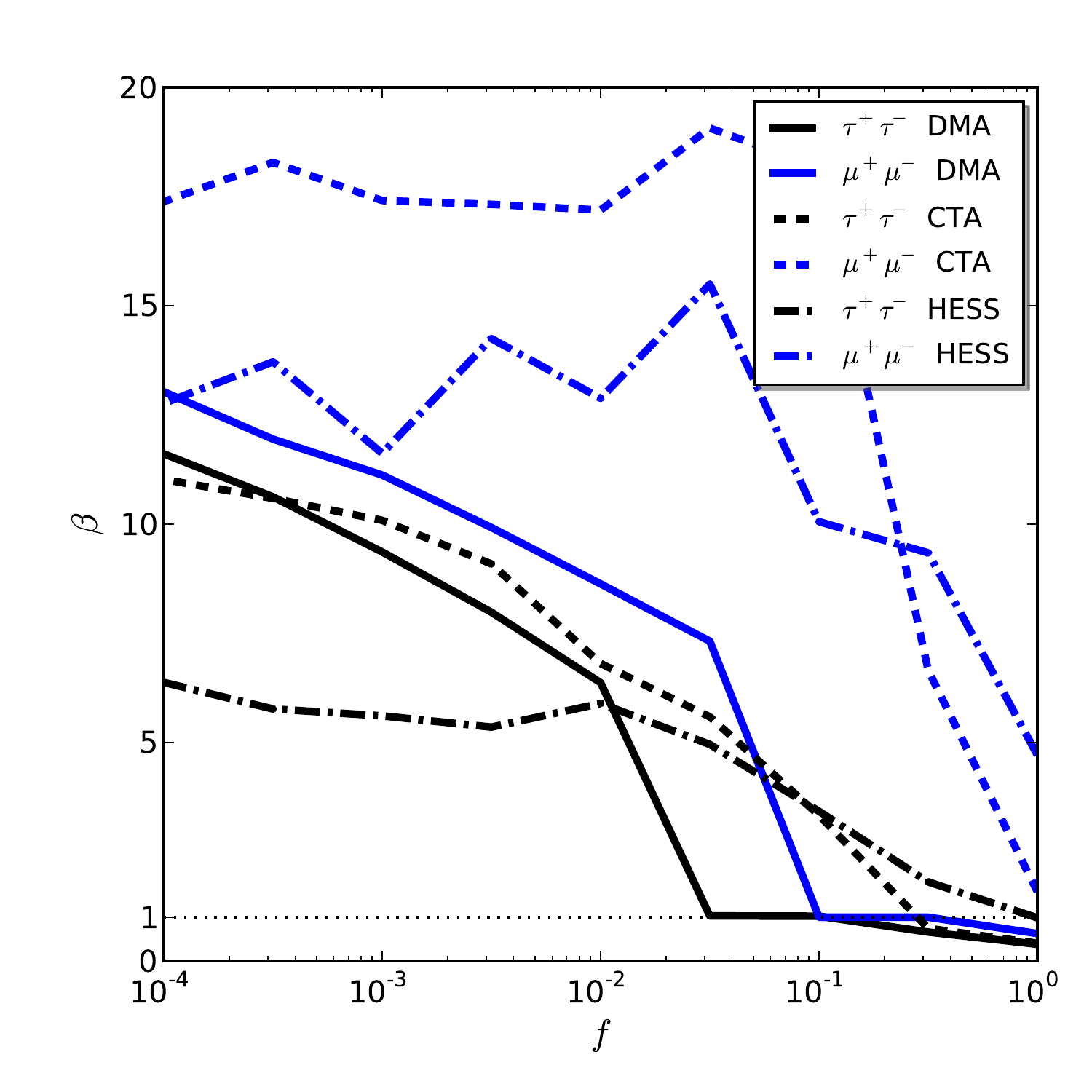} &
\includegraphics[width=0.3\linewidth,clip=]{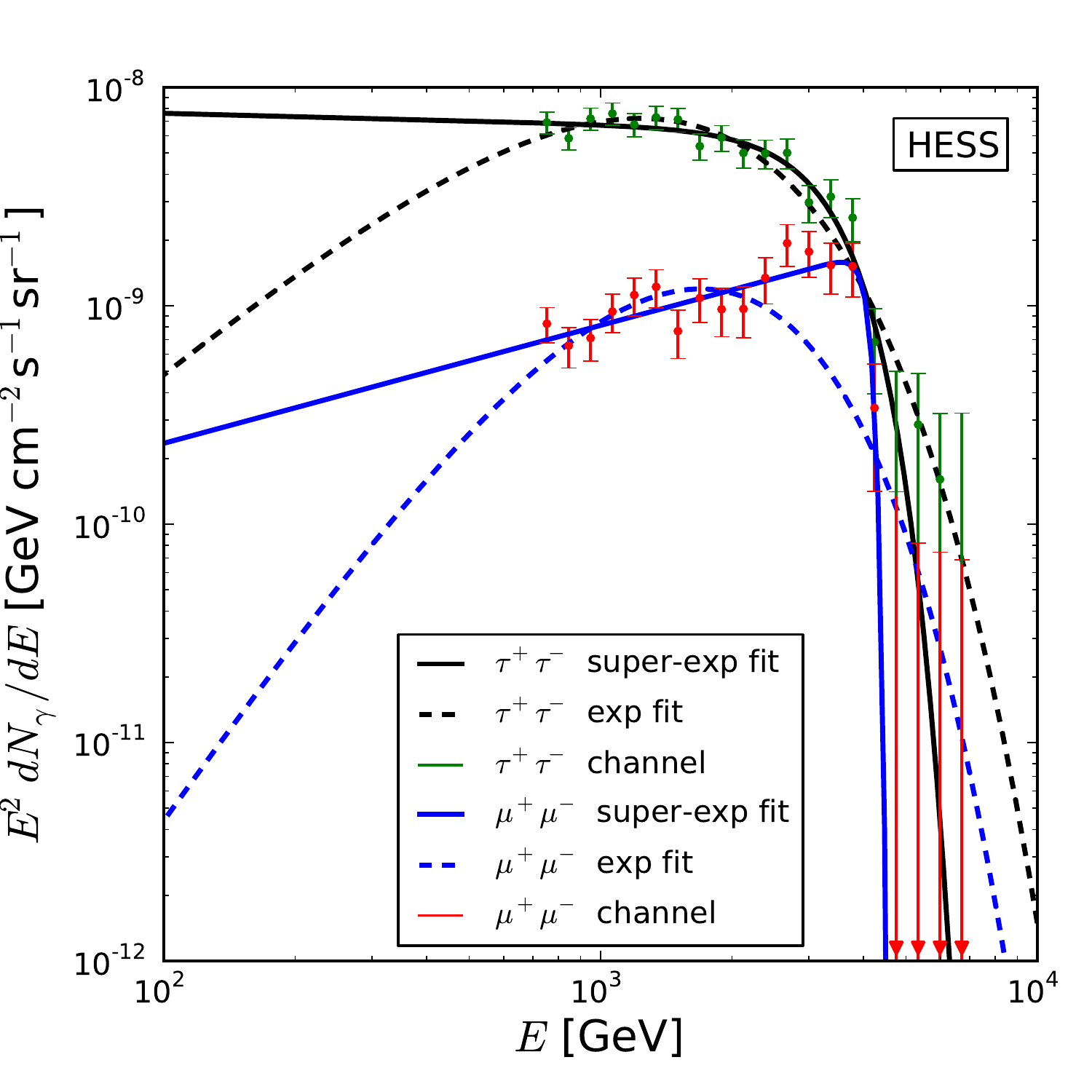} &
\includegraphics[width=0.3\linewidth,clip=]{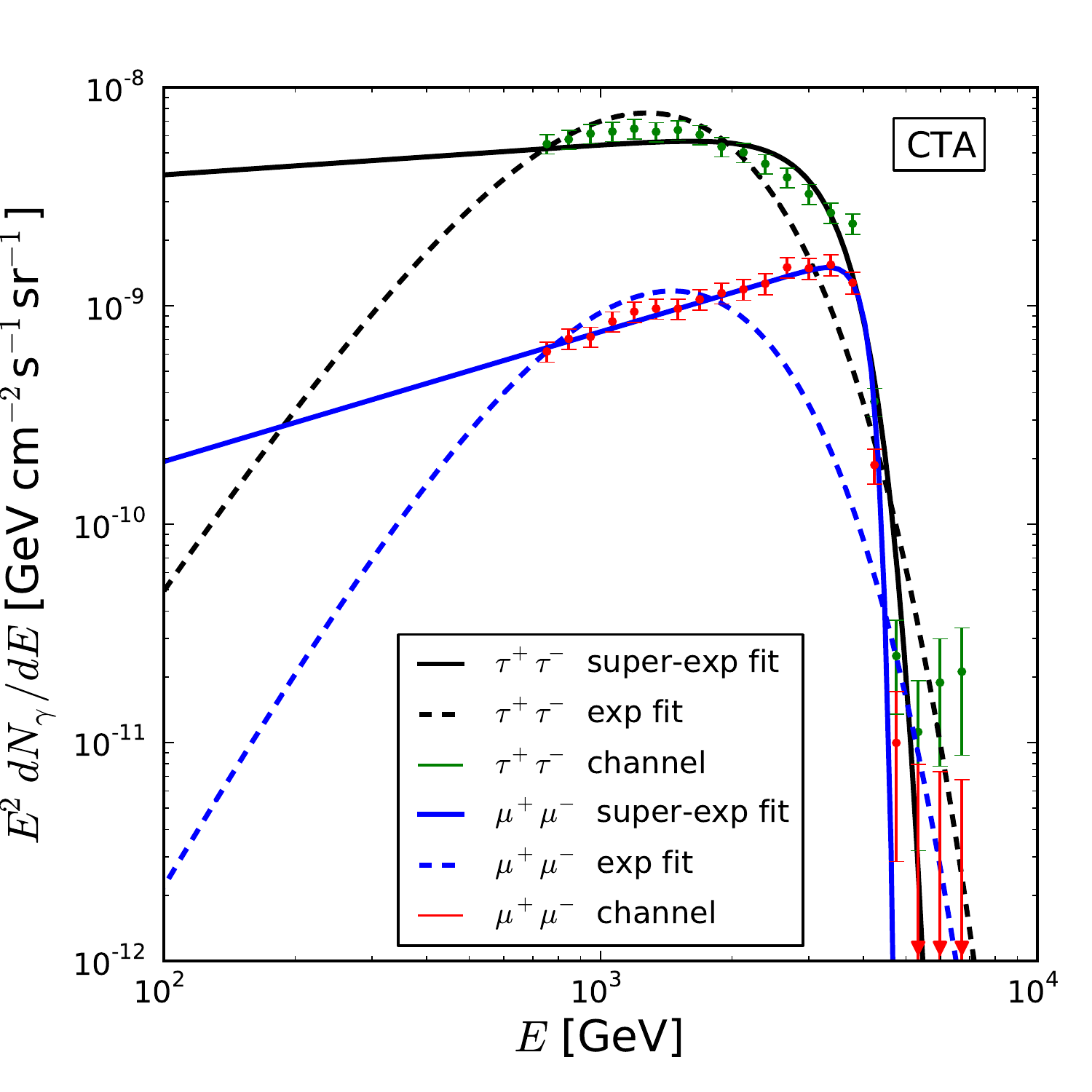}
\end{tabular}
\caption{
Averaged $\beta$ of best fits as a function of $f$, the fraction of energy in the background (left): HESS,
 dash-dotted lines; CTA, dashed lines, and DMA, solid lines. Blue lines (upper) correspond to $\mu^+ \mu^-$ annihilations and black lines (lower) to $\tau^+ \tau^-$. Here $\bar E = \varepsilon^{-0.25} M_\chi$, $\varepsilon= 10$.
Best spectral fits to $\tau^+ \tau^-$ and $\mu^+ \mu^-$ annihilation channels for HESS (middle) and CTA (right). Solid (super-exponential fit) and dashed (exponential fit) black lines (top) correspond to $\tau^+ \tau^-$ annihilations, and solid (super-exponential fit) and dashed (exponential fit) blue lines (bottom) to $\mu^+ \mu^-$.
}
\label{fig:BetaofAPLFits}
\end{figure*}

Having confirmed that for certain values of $f$ the flux observable by ground-based telescopes can be well fit for $\varepsilon \sim 10$, we study the goodness of fit as a function of $f$, the normalization of the hypothetical background (having fixed $\varepsilon = 10$).
In Fig. \ref{fig:chi2AsFunctionOfApl}, one can see that with the increase of $f$, the exponential fits become better, while the super-exponential fits become worse, until at $f \simeq 0.1$ they become comparable. Thus when the background flux becomes a large enough fraction of the foreground flux (and specifically becomes comparable around the cutoff energy), the relevance of sharp-featured templates is diminished. In Fig. \ref{fig:BetaofAPLFits}, left inset, the average exponents $\beta$ are plotted as functions of $f$. The average $\beta$ values of the best fits to the $\mu^+ \mu^-$ annihilation channel are systematically greater than those of the $\tau^+ \tau^-$ annihilation channel. We also note the average $\beta$ values of the best fits are systematically greater for fits to CTA over HESS experiments. 
Examples of best fits to particular realization of fluxes observed from the Galactic Center by HESS and CTA are presented in Fig. \ref{fig:BetaofAPLFits} (center and right).

In summary, we have investigated the use of super-exponential versus exponential templates for fitting the signal of the flux of gamma-rays from dark matter annihilating at the Galactic Center and an unknown astrophysical background simulated by a power-law. We find that 
the super-exponential template provides better fits than the exponential template 
with good $\chi^2 \lesssim 1$ for the HESS and the CTA experiments. 
If dark matter annihilates through $\mu^+ \mu^-$ or $\tau^+ \tau^-$ channels then the use of the super-exponential template is more robust relative to other channels provided that the ratio of signal to background fluxes is smaller than 0.1. This allows us to assert that if the background can be discriminated sufficiently well, the exponential fits to the combined gamma-ray flux from the Galactic Center are significantly worse than the super-exponential ones. In this case, the signal could be plausibly explained by dark matter annihilating to $\mu^+ \mu^-$ or $\tau^+ \tau^-$ annihilation channels. The inverse statement is even stronger - if there is no gain by using a super-exponential template over an exponential one, we can place strong constraints on $\mu^+ \mu^-$ or $\tau^+ \tau^-$ and weaker on other annihilation channels.

We thank T. Bringmann, E. Moulin, M. Cirelli for useful conversations and the referee for helpful comments. The authors were supported by ERC project 267117 (DARK MATTERS) hosted by Universite Pierre et Marie Curie - Paris 6 and acknowledge the hospitality of KICP where part of this work was done.

\end{document}